\documentclass[11pt,a4paper]{article}
\pdfoutput=1 

\usepackage[table]{xcolor}
\usepackage{colortbl}
\RequirePackage{ifpdf} 
\usepackage{amsmath} 
\usepackage{mathtools}

\usepackage{jheppub}
\usepackage{pstricks}
\usepackage[final]{pdfpages} 
\usepackage{ifpdf} 
\usepackage{slashed}

\usepackage[normalem]{ulem}
\usepackage{color}
\usepackage{xcolor}
\definecolor{urlblue}{rgb}{0.2,0.4,0.7}
\definecolor{citegreen}{rgb}{0,0.6,0.2}
\definecolor{linkred}{rgb}{0.9,0.2,0.1}
\usepackage{hyperref}
\hypersetup{
colorlinks=true, citecolor=citegreen, linkcolor=blue, urlcolor=urlblue}

\usepackage{graphics}
\usepackage{etoolbox} 
\usepackage{fixmath}
\usepackage{psfrag}

\usepackage{notoccite} 

\usepackage{amsfonts}
\usepackage{autobreak}
\usepackage{marginnote}
\usepackage{enumitem}
\usepackage{appendix}

\newcommand{\NOdisplay}[1]{ }

\def\MSbar{\overline{\mathrm{MS}}}

\def\diagen{{\fontfamily{qcr}\selectfont
DiaGen}}

\def\canonica{{\fontfamily{qcr}\selectfont
CANONICA}}

\title{The complete singlet contribution to the massless quark form factor at three loops in QCD}

\author{Long Chen$^{a}$, Micha\l{} Czakon$^{a}$, Marco Niggetiedt$^{a}$}
\emailAdd{longchen@physik.rwth-aachen.de, mczakon@physik.rwth-aachen.de,marco.niggetiedt@rwth-aachen.de}
\affiliation{
$^a$Institut f\"ur Theoretische Teilchenphysik und Kosmologie, RWTH Aachen University,\\ Sommerfeldstr.~16, 52056 Aachen, Germany}

\preprint{TTK-21-34,~P3H-21-059}

\abstract{
It is well known that the effect of top quark loop corrections in the axial part of quark form factors (FF) does not decouple in the large top mass or low energy limit due to the presence of the axial-anomaly type diagrams.
The top-loop induced singlet-type contribution should be included in addition to the purely massless result for quark FFs when applied to physics in the low energy region, both for the non-decoupling mass logarithms and for an appropriate renormalization scale dependence.
In this work, we have numerically computed the so-called singlet contribution to quark FFs with the exact top quark mass dependence over the full kinematic range.
We discuss in detail the renormalization formulae of the individual subsets of the singlet contribution to an axial quark FF with a particular flavor, as well as the renormalization group equations that govern their individual scale dependence.
Finally we have extracted the 3-loop Wilson coefficient in the low energy effective Lagrangian, renormalized in a non-$\MSbar$ scheme and constructed to encode the leading large mass approximation of our exact results for singlet quark FFs. 
We have also examined the accuracy of the approximation in the low energy region.
}

\begin{document}
\allowdisplaybreaks[4]
\unitlength1cm
\maketitle
\flushbottom

\section{Introduction}
\label{sec:intro}

The form factors (FF) of the vertices that couple an external color-neutral boson, such as a Higgs or an electroweak gauge boson, to a pair of quarks or gluons are important ingredients for calculating a number of phenomenologically interesting processes.
The knowledge of high order perturbative corrections to these vertex FFs in Quantum Chromodynamics (QCD) is essential to make precision predictions for collider processes such as quark pair production in electron-position collisions, the Drell-Yan processes, hadronic production and decay of the Higgs boson and massive electroweak bosons.
Furthermore, these vertex FFs constitute simple yet important objects of which the high-order QCD corrections can be used to extract certain universal QCD quantities of particular theoretical interest, such as the cusp anomalous dimensions~\cite{Korchemsky:1987wg,Korchemskaya:1992je} and the collinear quark and gluon anomalous dimensions (see, e.g.,~ref.~\cite{Moch:2005id,Becher:2006mr,Becher:2009qa,Henn:2016wlm,Lee:2017mip,Lee:2019zop,vonManteuffel:2020vjv}).

Due to the aforementioned importance, there has been a great amount of work on these objects in the literature.
In this article, we are concerned with the so-called \textit{singlet} type contribution to quark FFs describing the coupling of a pair of massless quarks to an external (axial) vector current to three loops in QCD including effects of a massive top quark.
The QCD virtual corrections to quark FFs can be conveniently divided into two classes depending on whether the external color-neutral boson couples directly to the external quarks.
Such a separation of QCD corrections is convenient for a number of practical reasons, such as allowing one to apply different $\gamma_5$ prescriptions~\cite{tHooft:1972tcz,Akyeampong:1973xi,Breitenlohner:1977hr,Bardeen:1972vi,Chanowitz:1979zu,Gottlieb:1979ix,Ovrut:1981ne,Espriu:1982bw,Buras:1989xd,Kreimer:1989ke,Korner:1991sx,Larin:1991tj,Larin:1993tq,Jegerlehner:2000dz,Moch:2015usa,Zerf:2019ynn} in the case of an axial current, as well as simplification of the calculation of loop integrals if the internal fermion loops are massive.
Limited to purely massless QCD corrections, the 3-loop results for vector~\cite{Moch:2005id,Baikov:2009bg,Lee:2010cga,Gehrmann:2010ue}, scalar~\cite{Gehrmann:2014vha} and pseudo-scalar~\cite{Ahmed:2015qpa} part of quark FFs were derived in the literature, and there was recently important progress towards the 4-loop corrections to the vector quark  FFs~\cite{Henn:2019swt,Lee:2019zop,vonManteuffel:2020vjv,Agarwal:2021zft}.\footnote{We note that massive quark FFs are known to 2-loop order in QCD~\cite{Bernreuther:2004ih,Bernreuther:2004th,Bernreuther:2005rw,Bernreuther:2005gw}, and partially at 3-loop order~\cite{Henn:2016tyf,Lee:2018nxa,Lee:2018rgs,Ablinger:2018yae,Blumlein:2019oas}.}
The three-loop singlet contribution to the axial part of quark FFs in purely massless QCD was determined only very recently in ref.~\cite{Gehrmann:2021ahy}.

However, for physical application of the result for the axial quark FF, such as for theoretical predictions of the Z-mediated Drell-Yan processes to the third order in QCD coupling $\alpha_s$, it is necessary to incorporate the singlet QCD contribution with top quark loops, for at least two reasons both related to the presence of the axial-anomaly type diagrams~\cite{Adler:1969gk,Bell:1969ts}. 
First, in the absence of the top-loop contribution, the purely massless contribution to the axial FFs contains an explicit logarithmic renormalization scale dependence beyond that expected from the running of the $\MSbar$ renormalized $\alpha_s$, which is related to the non-vanishing anomalous dimension of the singlet axial current (e.g.,~determined in refs.~\cite{Larin:1991tj,Larin:1993tq,Ahmed:2021spj}).
Second, as well known in refs.~\cite{Collins:1978wz,Chetyrkin:1993jm,Chetyrkin:1993ug,Larin:1993ju,Larin:1994va}, the top quark contribution to the axial FFs does not actually decouple in the large top mass or low energy limit, in contrast to the case of vector FFs.
In particular, the singlet-type QCD contribution to the inclusive Z boson decay rate has been investigated in detail in the large top mass limit to  $\mathcal{O}(\alpha_s^3)$~\cite{Chetyrkin:1993ug,Larin:1993ju,Larin:1994va} and to $\mathcal{O}(\alpha_s^4)$~\cite{Baikov:2012er}, and was found to be considerable.

Here we compute the exact top quark contribution to massless quark FFs to 3-loop order in QCD, especially for the axial part.
Most of the top mass dependent master integrals involved can be mapped to those in the 3-loop Higgs-gluon FF determined in ref.~\cite{Czakon:2020vql}, and the additional ones are computed analytically and verified numerically by the same technique through this work. 
With the complete singlet axial current renormalization constant determined to $\mathcal{O}(a_s^3)$ in ref.~\cite{Ahmed:2021spj}, including the non-$\MSbar$ finite piece, we are able to properly treat the individual subsets of singlet diagrams separated according to the flavor of the internal quark coupled to the Z boson, namely each separated flavor subset is mathematically consistent on its own.
Not only an interesting theoretical question on its own, the UV renormalization of the anomalous top-quark loop contribution to the axial quark form factor determines the structure of the non-decoupling mass logarithms as well. We will therefore discuss the relevant renormalization formula in detail later in the article.
We note that as long as one is only concerned with the anomaly-free sum of all singlet-type QCD diagrams from each electroweak doublet, e.g.,~in refs.~\cite{Chetyrkin:1993jm,Chetyrkin:1993ug,Larin:1993ju,Larin:1994va,Baikov:2012er}, it is not necessary to include the non-$\MSbar$ part in the renormalization of the (singlet) axial current.
The result presented here provides one of the missing ingredients needed to push the theoretical predictions of Z-mediated Drell-Yan processes to the third order in  $\alpha_s$, such as done recently for those mediated by a virtual photon~\cite{Duhr:2020seh,Chen:2021vtu} or a W boson~\cite{Duhr:2020sdp}.
~\\

The article is organized as follows.
In the next section, we introduce our conventions and notations for the quark FFs, and subsequently discuss the technicalities of their computation in section~\ref{sec:calc}. 
In section~\ref{sec:uvir} we discuss in detail the ultraviolet (UV) renormalization formulae for the individual subsets of singlet contributions, as well as the renormalization-group (RG) equations that govern the scale dependence of their finite remainders defined after subtraction of infrared (IR) divergences.  
We then present our exact numerical results for these finite remainders in section~\ref{sec:res}, and examine the quality of the large mass expansion results.
In section~\ref{sec:wcv} we extract the Wilson coefficient in front of the axial current of massless quarks in the low energy effective QCD, which can be conveniently used to approximate the leading behavior of the singlet contribution in the large top mass limit.
We conclude in section~\ref{sec:conc}.

\section{Preliminaries}
\label{sec:prel}

We consider the one-particle-irreducible corrections to the 3-point vertex function of an external (off-shell) Z boson and a pair of massless quarks of flavor $q$ with on-shell outgoing momenta $p_1$ and $p_2$, in QCD with $n_f = n_l + 1 = 6$ flavors and only the top quark kept massive.  
This vertex function admits the following Lorentz tensor decomposition
\begin{eqnarray}
\label{eq:FFdef}
\bar{u}(p_1)\, \mathrm{\Gamma}^{\mu} \, v(p_2) \, \delta_{ij} 
= \bar{u}(p_1)\,\big( v_q\, \mathit{F}^{V} \gamma^{\mu} \,+\, a_q\, \mathit{F}^{A}  \gamma^{\mu} \gamma_5 \big) )\,v(p_2) \, \delta_{ij}
\end{eqnarray}
where $\delta_{ij}$ denotes the color factor, and $v_q$ and $a_q$ are respectively the vector and axial vector couplings of the external quark $q$ to the Z boson.
In eq.(\ref{eq:FFdef}), we have used the fact that limited to gauge interactions, there are only two Lorentz structures, one parity-even and the other parity-odd, sandwiched between the two on-shell massless spinors which are linearly independent in 4 dimensions.
The Lorentz-invariant coefficients $\mathit{F}^{V}$ and $\mathit{F}^{V}$ are, respectively, the vector and axial FF of the massless quark $q$, which are functions of $s = (p_1 + p_2)^2 = 2\,p_1 \cdot p_2$ as well as the top mass $m_t$ (when the top quark loop contributes).   
The normalization is such that the tree-level values of these FFs read (in 4 dimensions): $\mathit{F}^{V,0} = 1\,, \mathit{F}^{A,0} = 1$.

The two Lorentz structures in eq.(\ref{eq:FFdef}) are orthogonal to each other. 
Pulling out the color factor and putting $v_q = a_q = 1$, the dimensionless $\mathit{F}^{V}$ and $\mathit{F}^{A}$ can be projected out in the following way:
\begin{eqnarray}
\label{eq:ProjV}
\mathit{F}^{V} &=& \frac{-1}{s (4 - 4 \epsilon)}\, \mathrm{Tr} \big[\slashed{p}_2 \gamma_{\mu} \,\slashed{p}_1 \, \mathrm{\Gamma}^{\mu}\big]\,, \nonumber\\
\mathit{F}^{A} &=& \frac{-1}{s (4 - 4 \epsilon)}\, \mathrm{Tr} \big[\slashed{p}_2 \gamma_{\mu} \gamma_5 \,\slashed{p}_1 \, \mathrm{\Gamma}^{\mu}\big]\,,
\end{eqnarray}
with the Dirac algebra and trace $\mathrm{Tr}$ done in $D=4 - 2 \epsilon$ dimensions, and the $\gamma_5$ treated as anticommuting in the second line.  
In our calculations of the singlet contribution to $\mathit{F}^{A}$, we used a non-anticommuting $\gamma_5$ definition~\cite{tHooft:1972tcz,Breitenlohner:1977hr} in the variant as prescribed in refs.~\cite{Larin:1991tj,Larin:1993tq}.  
Notice that the same projection applies, even though the form of the projection has been determined assuming an anti-commuting $\gamma_5$.
In fact, as long as one is only concerned with the finite remainders of these FFs in 4 dimensions, to be discussed in the following, one can set the $\epsilon$-parameter in these projectors to be 0 from the outset~\cite{Chen:2019wyb,Ahmed:2019udm}, which is what we actually did regarding the axial FF projector.
~\\

The QCD virtual corrections to $\mathit{F}^{V}$ and $\mathit{F}^{A}$ can be conveniently classified into two parts, the non-singlet and singlet part, 
\begin{eqnarray}
\mathit{F}^{V}  &=& \mathit{F}^{V}_{ns} + \mathit{F}^{V}_{s} = \mathit{F}^{V}_{ns} + \sum_{f} \frac{v_f}{v_q}\,\mathit{F}^{V}_{s,f} \,, \nonumber\\
\mathit{F}^{A} &=& \mathit{F}^{A}_{ns} + \mathit{F}^{A}_{s} = \mathit{F}^{A}_{ns} + \sum_{f} \frac{a_f}{a_q}\, \mathit{F}^{A}_{s,f} \,,
\end{eqnarray}
depending on whether the external Z boson couples directly to the external quarks or not. 
For the sake of later convenience, we have pulled out the Z boson couplings from the respective singlet contribution.
In the remainder of this article, we adopt the convention regarding the terminology for the singlet and non-singlet type QCD corrections to $\mathit{F}^{V(A)}$ where the classification is solely based on the topology of the contributing Feynman diagrams. 
The non-singlet QCD corrections have the Z boson coupled directly to the open fermion line of the external quark $q$, which starts from the tree level. 
It thus depends only on the electroweak coupling of the external quark $q$. 
With $q$ massless and an anticommuting $\gamma_5$ (which is straightforward to apply here), one has $\mathit{F}^{A}_{ns} = \mathit{F}^{V}_{ns}$ to all orders in QCD owing to chirality conservation.

\begin{figure}[htbp]
\begin{center}
\includegraphics[scale=0.45]{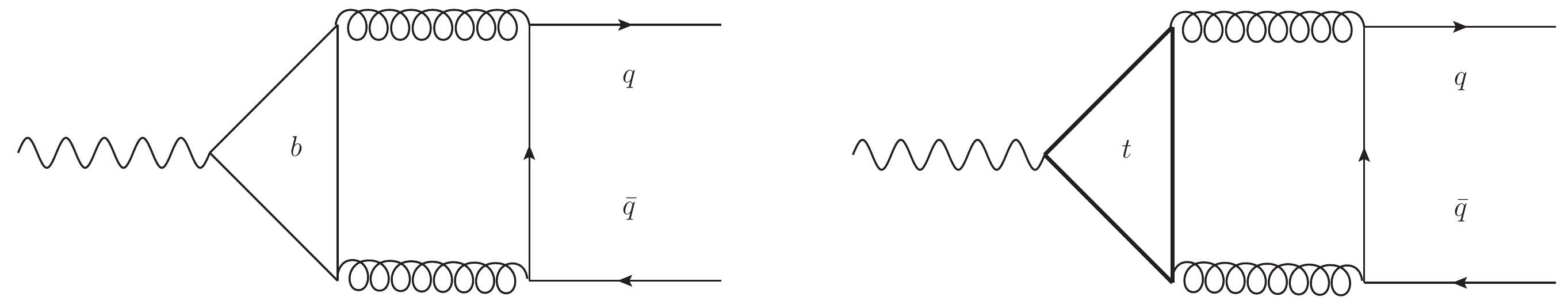}
\caption{Samples of singlet diagrams at 2-loop order.}
\label{fig:singlet2L}
\end{center}
\end{figure}

On the other hand, the singlet contribution $\mathit{F}^{V(A)}_{s}$ features a closed fermion loop which contains the quark coupling to the Z boson and starts in general from the 2-loop order as illustrated in fig.~\ref{fig:singlet2L}.
Consequently, $\mathit{F}^{V(A)}_{s}$ is associated with the electroweak couplings of the internal quarks running in the loops, which are normalized w.r.t that of external quark $q$ as defined in eq.(\ref{eq:FFdef}).
In the Standard Model, quarks in a weak doublet couple with opposite sign to the Z boson in the axial part of the neutral current, and hence axial contributions from doublets add up to zero in the massless limit for singlet diagrams.  
Therefore in the usual approximation taken here, the only non-zero axial contribution comes from the top-bottom doublet due to the large mass difference.
Specifically,  we denote 
\begin{eqnarray}
\label{eq:totSFF}
\mathit{F}^{A}_{s}  &=&\lambda_q \, \big(\mathit{F}^{A}_{s,b} - \mathit{F}^{A}_{s,t}\big)\,,
\end{eqnarray} 
with $\lambda_q \equiv \frac{a_b}{a_q}$ equal to $\pm 1$ depending on whether the external $q$ is an upper or lower quark (i.e.,~having the same weak isospin as the bottom quark). 
The full QCD corrections to $\mathit{F}^{A}_{s}$ were determined to 2-loop order in refs.~\cite{Bernreuther:2005rw} for both massless and massive external quarks.   
Very recently the 3-loop bottom contribution $\mathit{F}^{A}_{s,b}$ was derived in effective QCD with $n_l$ massless quarks in ref.~\cite{Gehrmann:2021ahy}.
In this work, we provide the result for the top-loop contribution $\mathit{F}^{A}_{s,t}$ to 3-loop order with exact top mass dependence, and also the part $\mathit{F}^{A}_{s,b}$ that contains top quark loops inserted through gluon self-energy corrections.

\begin{figure}[htbp]
\begin{center}
\includegraphics[scale=0.45]{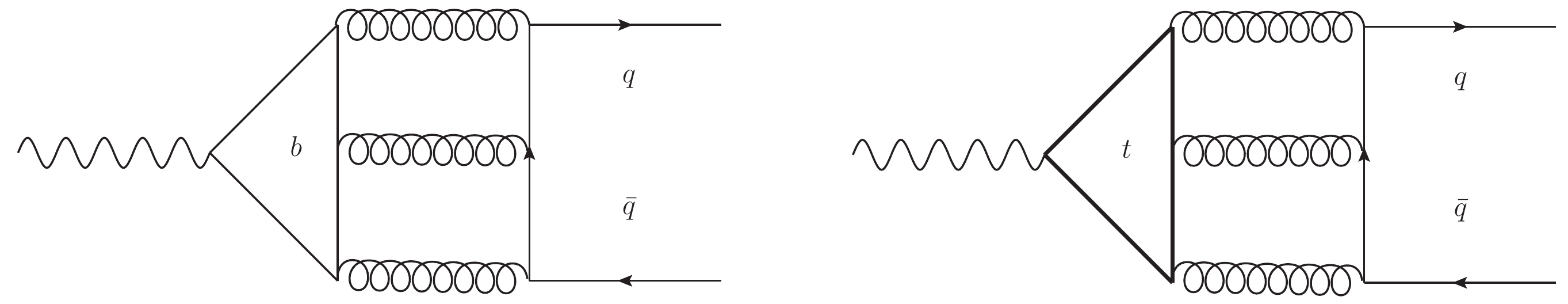}
\caption{Examples of singlet diagrams contributing to the vector part of quark form factors.}
\label{fig:singlet3L_VFF}
\end{center}
\end{figure}

Concerning the vector part of the singlet contribution, $\mathit{F}^{V}_{s}$ vanishes at the 2-loop order due to the same reason that underlies the Furry theorem, and starts to contribute only from the 3-loop order with sample diagrams shown in fig.~\ref{fig:singlet3L_VFF}. 
The leading 3-loop result is completely UV and IR finite, as computed in refs.~\cite{Moch:2005id,Baikov:2009bg,Gehrmann:2010ue} but with only massless quarks included.
The previously-missing top-loop induced contribution, i.e.,~the diagram with thick lines in fig.~\ref{fig:singlet3L_VFF}, is computed in this work for completeness. 
It is known to be power suppressed in the low energy limit.

\section{Calculation of the bare singlet form factors}
\label{sec:calc}

The bare quark FFs can be expanded formally in the bare QCD coupling constant $\hat{a}_s \equiv \frac{\hat{\alpha}_s}{4 \pi}$:  
\begin{eqnarray}
\label{eq:FFexpand}
\mathit{F}^{A}_{s,b} = \sum_{n = 2}^{\infty} \hat{a}^n_s \, \mathit{F}^{A, n}_{s,b} \,, \quad \, 
\mathit{F}^{A}_{s,t} = \sum_{n = 2}^{\infty} \hat{a}^{n}_s \, \mathit{F}^{A, n}_{s,t} \,,
\end{eqnarray}
where the perturbative expansion starts from the 2-loop order.
For the calculation of these QCD loop corrections, we use dimensional regularization~\cite{tHooft:1972tcz,Bollini:1972ui} with a non-anticommuting $\gamma_5$~\cite{tHooft:1972tcz,Breitenlohner:1977hr} in a variant as prescribed in refs.~\cite{Larin:1991tj,Larin:1993tq}.
The techniques employed closely follow those used in the computation of Higgs-gluon FF in ref.~\cite{Czakon:2020vql}. 
Here we merely sketch the main points specific to the amplitude in question.
Symbolic expressions of the contributing Feynman diagrams to the 3-loop order are generated by the C++ diagram generator \diagen~\cite{diagen}. 
There are 2 at the 2-loop order, and 57 at the 3-loop order, for both $\mathit{F}^{A}_{s,b}$ and $\mathit{F}^{A}_{s,t}$.
There are 4 diagrams contributing to $\mathit{F}^{A,3}_{s,b}$ with top-quark loops (see fig.~\ref{fig:singlet3L_topbub} for an example), while the remaining ones are completely massless and have been determined in ref.~\cite{Gehrmann:2021ahy}, which we include here only for completeness.

The loop integrals in all contributing diagrams are then reduced to a much smaller set of master integrals via Integration-By-Parts (IBP) identities~\cite{Tkachov:1981wb,Chetyrkin:1981qh} with the help of a C++ implementation of the Laporta algorithm~\cite{Laporta:2001dd}.
All massless master integrals involved in $\mathit{F}^{A}_{s,b}$ are known analytically~\cite{Moch:2005id,Baikov:2009bg,Lee:2010cga,Gehrmann:2010ue}.
The master integrals in $\mathit{F}^{A}_{s,t}$ can all be mapped to those solved numerically in ref.~\cite{Czakon:2020vql}.
There are, however, 15 top mass dependent masters $M_{i}(\epsilon, \frac{s}{m_t^2})$ (with $i = 1,2,\cdots, 15$) appearing in $\mathit{F}^{A,3}_{s,b}$, which we have to solve in addition in the present work.
They are, fortunately, simple enough to be solved analytically using the differential equation approach~\cite{Kotikov:1991pm,Remiddi:1997ny}.
We first derive the system of first-order homogeneous linear differential equations in the variable $ x\equiv \frac{s}{m_t^2}$ for these 15 masters:  
\begin{eqnarray}
\label{eq:MIsDE}
\frac{\mathrm{d}\, M_{i}(\epsilon, x)}{\mathrm{d}\, x} 
= \sum_{j} A_{ij} (\epsilon, x)\, M_{j}(\epsilon, x)\,,
\end{eqnarray}
by IBP reducing their derivatives back to themselves.
The coefficients $A_{ij} (\epsilon, x)$ are rational functions in $x$ and $\epsilon$ by the virtue of IBP identities.
After performing a change of variable~\footnote{We note that the variable y should have a positive imaginary part for $0 < x < 4$.} $x = 2 - y - \frac{1}{y}$, the differential equation system (\ref{eq:MIsDE}) is then fed to the package~\canonica~\cite{Meyer:2016slj,Meyer:2017joq}, which finds an $\epsilon$-form~\cite{Henn:2013pwa} together with the rational transformation of the basis of master integrals\footnote{We note that the basis found by \canonica~for this system is not of uniform weight, and thus does not really qualify as a usual \textit{canonical} basis~\cite{Henn:2013pwa}.}. 
The \textit{letters} involved in this $\epsilon$-form differential equation are $\{y,\, y+1,\, y-1\}$, and it is then straightforward to read off the solutions in terms of Harmonic polylogarithms (HPL)~\cite{Remiddi:1999ew} with certain integration constants to be fixed by boundary conditions.
We determine the boundary conditions by computing these master integrals in the large mass limit, which is located at $x \rightarrow 0$ (or $y \rightarrow 1$).
In this way, all 15 top mass dependent masters in $\mathit{F}^{A,3}_{s,b}$ are solved analytically in terms of HPLs, expanded in $\epsilon$ to the orders needed to obtain the $\mathit{F}^{A,3}_{s,b}$ at $\mathcal{O}(\epsilon^0)$.
In the supplemental material associated with this article, we attach the analytical result for the UV renormalized top-quark loop contribution to $\mathit{F}^{A,3}_{s,b}$.

\section{The renormalization formulae and RG equations}
\label{sec:uvir}

With the bare results for $\mathit{F}^{A}_{s,b}$ and $\mathit{F}^{A}_{s,t}$ at hand, we are now ready to perform the UV renormalization and define the finite remainders after IR subtraction.
Although the non-singlet and singlet part of the axial quark FF contribute to physical observables in a coherent physically-indistinguishable way, they do depend on (potentially) different electroweak couplings. 
Therefore, as far as the QCD corrections are concerned, they can be treated independently. 
In particular, one can derive RG equations for them separately.

To set up the notations and conventions in use, let us start with the renormalization of the bare QCD coupling ${\hat a}_s$,  
\begin{eqnarray}
\label{eq:asuvr}
{\hat a}_s \, \,S_{\epsilon} = Z_{a_s}(\mu^2)\, a_s(\mu^2) \, \mu^{2\epsilon}\,,
\end{eqnarray}
with the dimensionless renormalized coupling $a_s \equiv \frac{\alpha_s}{4 \pi} = \frac{g_s^2}{16 \pi^2}$.
The bare coupling ${\hat a}_s$ has mass-dimension $2\epsilon$ as is exhibited on the r.h.s. of \eqref{eq:asuvr}.
We work in the $\MSbar$ scheme for dimensionally-regularized loop integrals, $S_{\epsilon} = \left(4 \pi \right)^{\epsilon} e^{-\epsilon \gamma_E}$ (with $\gamma_E$ the Euler constant).
All UV poles on the r.h.s.~of eq.~\eqref{eq:asuvr} are explicitly encoded in $Z_{a_s}$ whose dependence on the scale $\mu$ is implicit and enters solely through the renormalized coupling $a_s$. (The dependence on $\mu$ of these quantities is suppressed from here onward whenever there is no confusion.)
The independence of ${\hat a}_s$ on $\mu$ implies the RG equation of $a_s$ in $D$ dimensions:
\begin{eqnarray}
\label{eq:beta}
\mu^2\frac{\mathrm{d} \ln a_s}{\mathrm{d} \mu^2} = -\epsilon - \mu^2\frac{\mathrm{d} \ln Z_{a_s}}{\mathrm{d} \mu^2} \equiv -\epsilon + \beta \,,
\end{eqnarray}
where $\beta \equiv - \mu^2\frac{\mathrm{d} \ln Z_{a_s}}{\mathrm{d} \mu^2}$ denotes the QCD beta function, i.e.,~the anomalous dimension of the renormalized $a_s$ in 4 dimensions.
When discussing the extraction of the Wilson coefficient in the low energy effective theory in section~\ref{sec:wcv}, we will further perform an additional finite renormalization of $a_s$ to decouple the top quark effect in the gluon self-energy correction.

By the virtue of the multiplicative renormalizablity of the QCD Lagrangian and the definition of the renormalized axial currents with a non-anticommuting $\gamma_5$ as summarized in ref.~\cite{Ahmed:2021spj}, we derive the following renormalization formulae for the $a_b$- and $a_t$-dependent singlet contributions individually:
\begin{eqnarray}
\label{eq:UVsFF}
\mathbf{F}^{A}_{s,b}(a_s, m_t,\mu)
&=& Z_{ns}\, Z_2\, \mathit{F}^{A}_{s,b}(\hat{a}_s, \hat{m}_t) \,+\, Z_s\, Z_2 \Big( \mathit{F}^{A}_{ns}(\hat{a}_s, \hat{m}_t) + \sum_{i=1}^{n_f}\mathit{F}^{A}_{s,i}(\hat{a}_s, \hat{m}_t)  \Big) \,, \nonumber\\
\mathbf{F}^{A}_{s,t}(a_s, m_t,\mu)
&=& Z_{ns}\, Z_2\, \mathit{F}^{A}_{s,t}(\hat{a}_s, \hat{m}_t) \,+\, Z_s\, Z_2 \Big( \mathit{F}^{A}_{ns}(\hat{a}_s, \hat{m}_t) + \sum_{i=1}^{n_f}\mathit{F}^{A}_{s,i}(\hat{a}_s, \hat{m}_t)  \Big) \,,
\end{eqnarray} 
where the $\hat{a}_s$ on the r.h.s will be substituted according to eq.(\ref{eq:asuvr}) and the bare mass $\hat{m}_t$ is to be renormalized on-shell by $\hat{m}_t = Z_m\, m_t$. 
The dependence on $s$ is suppressed in order not to overload the notations.
It is understood that $\sum_{i=1}^{n_f}$ is a shorthand notation for $\sum_{{\tiny i \in\{u,\cdots,t\}}}$, and similarly $\sum_{i=1}^{n_l}$ (to appear below) refers to  $\sum_{{\tiny i \in\{u,\cdots,b\}}}$.
The $Z_2$ is the on-shell wavefunction renormalization constant of the external light quark, which differs from one due to the presence of massive top loops starting from 2-loop order. 
The $Z_s \equiv \frac{1}{n_f} \big( Z_S - Z_{ns} \big)$ is the difference between the usual \textit{singlet} and non-singlet axial current renormalization constants determined in refs.~\cite{Larin:1991tj,Ahmed:2021spj}, further normalized to the case of having just one single flavor, e.g.,~either the bottom or top quark, coupled to the Z boson. 
To be more specific, the constant $Z_{S}$ is defined in the following renormalized \textit{singlet} axial current
\begin{eqnarray}
\label{eq:ZScurrent}
\big[J^{\mu}_{S,5}\big]_R = Z_S \, J^{\mu}_{S,5} = Z_S \,\sum_{i=1}^{n_f} \, \bar{\psi}^{B}_{i}  \, \gamma^{\mu}\gamma_5 \, \psi^{B}_i\,,
\end{eqnarray}
with a non-anticommuting $\gamma_5$, and is given by the product of eq.(5.1) and eq.(5.4) of ref.~\cite{Ahmed:2021spj}. 
And to distinguish the different definitions of the terminology ``singlet'' we have given it a capitalized subscript.
The constant $Z_{ns}$ is the one needed to renormalize the \textit{non-singlet} axial current $J^{\mu}_{ns,5} = \sum_{i=1}^{n_f=6} a_i \, \bar{\psi}^{B}_{i}  \, \gamma^{\mu}\gamma_5 \, \psi^{B}_i$ with a non-anticommuting $\gamma_5$ 
and can be obtained by taking the product of eq.(8) and eq.(11) of ref.~\cite{Larin:1991tj}.
For reader's convenience, we reproduce here the result for $Z_s$, which reads 
\begin{eqnarray}
\label{eq:Zs}
Z_s &=& \,  a_s^2 \,C_F\, \Big(\, \frac{3}{\epsilon} \,+\, \frac{3}{2}\, \Big) \nonumber\\
&+& \, a_s^3\, \Big(\,  
C_A\,C_F\,  
\Big( -\frac{22}{3} \frac{1}{\epsilon^2} + \frac{109}{9} \frac{1}{\epsilon} - \frac{163}{27} + 26\, \zeta_3 \Big) \nonumber\\
&+&\, C_F^2\,  \Big(
-\frac{18}{\epsilon} + \frac{23}{2} - 24 \, \zeta_3 
 \Big) \, + \, C_F\,n_f\, \Big(\,
\frac{4}{3}\frac{1}{\epsilon^2}+\frac{2}{9}\frac{1}{\epsilon} + \frac{88}{27}  \Big)\, 
\Big)\, +\, \mathcal{O}(a_s^4)\,. 
\end{eqnarray}
The definition of the quadratic Casimir color constants is as usual: $C_A = N_c \,, \, C_F = (N_c^2 - 1)/(2 N_c)$ with the number of colors in the fundamental representation $N_c=3$ in QCD.
Based on the definition of $Z_s$, one has 
\begin{eqnarray}
\label{eq:ZsAMD}
\mu^2\frac{\mathrm{d} Z_s}{\mathrm{d} \mu^2}  = \frac{1}{n_f} \gamma_S \, Z_S \equiv \gamma_s \big(Z_{ns} + n_f \, Z_{s}\big) 
\end{eqnarray}
with $\gamma_S \equiv n_f \, \gamma_s =  \mu^2\frac{\mathrm{d} \ln Z_S}{\mathrm{d} \mu^2}$.
Notice that $Z_{ns}$ is scale independent, since the renormalized non-singlet axial and vector current are related by multiplication of $\gamma_5$ and the vector current is not renormalized.
It is worthy to emphasize that what appears in the r.h.s of eq.~(\ref{eq:ZsAMD}) is $Z_{ns} + n_f \, Z_{s}$ rather than just $Z_{s}$.
Furthermore, the $\mathit{F}^{A}_{ns}(\hat{a}_s, \hat{m}_t)$ in eq.(\ref{eq:UVsFF}) must be computed using the same non-anticommuting $\gamma_5$ prescription as used in calculating the singlet FFs. 
Therefore, this renormalization formula shows that as soon as one applies a non-anticommuting $\gamma_5$ prescription in the calculation of an isolated individual subset of singlet diagrams, one is forced to have the knowledge of the purely non-singlet diagrams in the same convention for the sake of renormalization, albeit only up to an order less by two loops.

Note that our formula eq.(\ref{eq:UVsFF}) is more general than what is needed here: once expanded up to $\mathcal{O}(a^3_s)$ in question, only
\begin{center}
$Z_{ns}\, Z_2\, \mathit{F}^{A}_{s,b}(\hat{a}_s, \hat{m}_t) \,+\, Z_s\, Z_2\, \mathit{F}^{A}_{ns}(\hat{a}_s, \hat{m}_t)$ 
\end{center}
actually contribute, because the singlet quantities, $Z_s$ and $\mathit{F}^{A}_{s}$, all start from $\mathcal{O}(a^2_s)$. 
For the same reason, the on-shell $Z_2$ does not contribute neither in our 3-loop results.
The remaining terms, which are quite non-trivial due to mixing with singlet diagrams featuring quarks of other flavors (with potentially different masses), should get involved but only starting from 4-loop order.
We note also that the appearance of the $\hat{m}_t$ dependence in various pieces in eq.(\ref{eq:UVsFF}) starts typically from 2-loop order, as the external quark $q$ is massless.
Furthermore, the difference between the two subsets of singlet contributions in eq.(\ref{eq:UVsFF}), which is essentially eq.(\ref{eq:totSFF}), requires only the non-singlet axial current renormalization:
\begin{eqnarray}
\label{eq:UVsFFdiff}
\mathbf{F}^{A}_{s,b}(a_s, m_t) - \mathbf{F}^{A}_{s,t}(a_s, m_t)
&=& Z_{ns}\, Z_2\, \big( \mathit{F}^{A}_{s,b}(\hat{a}_s, \hat{m}_t) \,-\, \mathit{F}^{A}_{s,t}(\hat{a}_s, \hat{m}_t) \big)\,,
\end{eqnarray} 
as expected.

In the case of the renormalization of the singlet contribution to the vector counterpart,  $\mathit{F}^{V}_{s,f}$, it is well known that one needs only the overall (on-shell) wavefunction renormalization constant $Z_2$ in addition to the renormalization of $\hat{a}_s$ and top mass. 
Furthermore, up to the $\mathcal{O}(a^3_s)$ considered in the present work, it is completely UV (and IR) finite.
~\\

Starting with eq.(\ref{eq:UVsFF}) and noting the non-zero anomalous dimension of the singlet axial-current operator eq.(\ref{eq:ZsAMD}), one can then derive the RG equations for the renormalized $a_b$- and $a_t$-dependent singlet contribution, which governs their respective $\mu$-dependence.
With the external quark field and top mass renormalized on-shell while the $a_s$ in the $\MSbar$ scheme, the RG equations read
\begin{eqnarray}
\label{eq:UVsFFsRGE}
\mu^2\frac{\mathrm{d}}{\mathrm{d} \mu^2} \mathbf{F}^{A}_{s,b}(a_s, m_t,\mu)
&=& \mu^2\frac{\partial}{\partial \mu^2} \mathbf{F}^{A}_{s,b}(a_s, m_t,\mu) \,+\, \big(\beta-\epsilon\big) a_s\frac{\partial}{\partial a_s} \mathbf{F}^{A}_{s,b}(a_s, m_t,\mu) \nonumber\\
&=& \gamma_s \Big( \mathbf{F}^{A}_{ns}(a_s, m_t,\mu) + \sum_{i=1}^{n_f}\mathbf{F}^{A}_{s,i}(a_s, m_t,\mu)  \Big) \,, \nonumber\\
\mu^2\frac{\mathrm{d}}{\mathrm{d} \mu^2} \mathbf{F}^{A}_{s,t}(a_s, m_t,\mu) 
&=& \mu^2\frac{\partial}{\partial \mu^2} \mathbf{F}^{A}_{s,t}(a_s, m_t,\mu) \,+\, \big(\beta-\epsilon\big) a_s\frac{\partial}{\partial a_s} \mathbf{F}^{A}_{s,t}(a_s, m_t,\mu) \nonumber\\
&=& \gamma_s \Big( \mathbf{F}^{A}_{ns}(a_s, m_t,\mu) + \sum_{i=1}^{n_f}\mathbf{F}^{A}_{s,i}(a_s, m_t,\mu)  \Big) \,,
\end{eqnarray} 
where all FFs on both sides are the UV renormalized ones. 
Again once expanded and truncated up to $\mathcal{O}(a^3_s)$, only the term with $\mathbf{F}^{A}_{ns}(a_s, m_t,\mu)$ in the r.h.s.  contributes to the 3-loop calculations considered in the present work.  
Just like the pure non-singlet contribution $\mathbf{F}^{A}_{ns}(a_s, m_t,\mu)$, one sees that the ``physical'' combination $\mathbf{F}^{A}_{s,b}(a_s, m_t,\mu) - \mathbf{F}^{A}_{s,t}(a_s, m_t,\mu)$ has a zero anomalous dimension as a direct consequence of eq.(\ref{eq:UVsFFdiff}):
\begin{eqnarray}
\label{eq:FFsRGEphys}
\mu^2\frac{\mathrm{d}}{\mathrm{d} \mu^2} 
\Big( \mathbf{F}^{A}_{s,b}(a_s, m_t,\mu) - \mathbf{F}^{A}_{s,t}(a_s, m_t,\mu) \Big)
&=& 0\,,  
\end{eqnarray}
which is also clear from eq.(\ref{eq:UVsFFsRGE}). 
Therefore, the net $\mu$ dependence in the $a_t$-dependent singlet contribution $\mathbf{F}^{A}_{s,t}(a_s, m_t,\mu)$ is necessary to cancel that of $\mathbf{F}^{A}_{s,b}(a_s, m_t,\mu)$, such that the remaining explicit $\mu$ dependence is related to the $\MSbar$ renormalization of $a_s$ in the usual way.  
Based on this, one anticipates already that the top-quark contribution $\mathbf{F}^{A}_{s,t}(a_s, m_t,\mu)$ cannot completely decouple in the naive sense in the large top mass or low energy limit, because the ``massless'' contribution $\mathbf{F}^{A}_{s,b}(a_s, m_t,\mu)$ still has a non-zero anomalous dimension to be compensated. 
~\\

The UV-renormalized singlet FFs still contain IR divergences, starting from 3-loop order, from exchange of virtual soft and/or collinear particles, regularized as poles in the dimensional regulator $\epsilon$. 
By factorizing out the IR singularities, we define the following finite remainders of $\mathbf{F}^{A}_{s,b(t)}(a_s, m_t,\mu)$:
\begin{eqnarray}
\label{eq:FiniteRemainder}
\mathcal{F}^{A}_{s,b}(a_s, m_t,\mu)
&=& I_{q\bar{q}}\, \mathbf{F}^{A}_{s,b}(a_s, m_t,\mu)  \nonumber\\
&=& a_s^2\, \mathcal{F}^{A,2}_{s,b}(\mu) \,+\, a_s^3\, \mathcal{F}^{A,3}_{s,b}(m_t,\mu) \,+\, \mathcal{O}(a_s^4) 
\,, \nonumber\\
\mathcal{F}^{A}_{s,t}(a_s, m_t,\mu)
&=& I_{q\bar{q}}\, \mathbf{F}^{A}_{s,t}(a_s, m_t,\mu) \nonumber\\
&=& a_s^2\, \mathcal{F}^{A,2}_{s,t}(m_t, \mu) \,+\, a_s^3\, \mathcal{F}^{A,3}_{s,t}(m_t,\mu) \,+\, \mathcal{O}(a_s^4) 
 \,,
\end{eqnarray} 
where the dependence on $s$ is suppressed as before, and the $I_{q\bar{q}}$ denotes the IR-singular factor determined in ref.~\cite{Catani:1998bh}. 
For the present application, the $I_{q\bar{q}}$ needed reads 
\begin{eqnarray}
\label{eq:Iqq}
I_{q\bar{q}} = 1 - 2\,a_s\, \left(\frac{\mu^2}{-s-i0^+}\right)^{\epsilon}\frac{e^{\epsilon \gamma_E}}{\Gamma(1-\epsilon)}\,C_F\,\left(\frac{1}{\epsilon^2} \,+\, \frac{3}{2\epsilon} \right) \,+\,\mathcal{O}(a_s^2).
\end{eqnarray}
With this $I_{q\bar{q}}$ operator, which has a vanishing anomalous dimension, the same form of the RG equations derived in eq.(\ref{eq:UVsFFsRGE}) simply carries over to the cases of finite remainders defined above, namely,
\begin{eqnarray}
\label{eq:FRsRGE}
\mu^2\frac{\mathrm{d}}{\mathrm{d} \mu^2} \mathcal{F}^{A}_{s,b}(a_s, m_t,\mu)
&=& \mu^2\frac{\partial}{\partial \mu^2} \mathcal{F}^{A}_{s,b}(a_s, m_t,\mu) \,+\, \beta\, a_s\frac{\partial}{\partial a_s} \mathcal{F}^{A}_{s,b}(a_s, m_t,\mu) \nonumber\\
&=& \gamma_s \Big( \mathcal{F}^{A}_{ns}(a_s, m_t,\mu) + \sum_{i=1}^{n_f}\mathcal{F}^{A}_{s,i}(a_s, m_t,\mu)  \Big) \,, \nonumber\\
\mu^2\frac{\mathrm{d}}{\mathrm{d} \mu^2} \mathcal{F}^{A}_{s,t}(a_s, m_t,\mu) 
&=& \mu^2\frac{\partial}{\partial \mu^2} \mathcal{F}^{A}_{s,t}(a_s, m_t,\mu) \,+\, \beta\, a_s\frac{\partial}{\partial a_s} \mathcal{F}^{A}_{s,t}(a_s, m_t,\mu) \nonumber\\
&=& \gamma_s \Big( \mathcal{F}^{A}_{ns}(a_s, m_t,\mu) + \sum_{i=1}^{n_f}\mathcal{F}^{A}_{s,i}(a_s, m_t,\mu)  \Big) \,,
\end{eqnarray}
where the $\mathcal{F}^{A}_{ns}$ and $\mathcal{F}^{A}_{s,i}$ denote the finite remainders of the corresponding UV-renormalized FFs defined in the same way as in eq.(\ref{eq:FiniteRemainder}), in particular by the same $I_{q\bar{q}}$ operator.
Once again, the combination $\mathcal{F}^{A}_{s}(a_s, m_t,\mu)\equiv\mathcal{F}^{A}_{s,b}(a_s, m_t,\mu) -  \mathcal{F}^{A}_{s,t}(a_s, m_t,\mu) $ has no anomalous dimension, just like the pure non-singlet contribution $\mathcal{F}^{A}_{ns}(a_s, m_t,\mu)$.
The top mass dependence in various parts above typically enter starting from $\mathcal{O}(a^2_s)$, except in $\mathit{F}^{A}_{s,b}$ where the dependence starts from the 3-loop order (see, e.g.~fig.~\ref{fig:singlet3L_topbub}). 
Since all IR-subtracted FFs involved in eq.(\ref{eq:FRsRGE}) are finite in 4 dimensions, one can simply take the 4-dimensional expressions of the anomalous dimensions involved, in particular the $\gamma_s$ as defined in eq.~(\ref{eq:ZsAMD}).

Alternatively, one could also consider finite remainders defined in the $\MSbar$ factorization scheme~\cite{Becher:2009cu} where the IR factors contain only poles.
Denoting the $\MSbar$ IR factor in need as $Z_{q\bar{q}}$, and by $\mathcal{F}^{'A}_{s,b}(a_s, m_t,\mu)$ the $\MSbar$ counterpart of  $\mathcal{F}^{A}_{s,b}(a_s, m_t,\mu)$, one has 
\begin{eqnarray}
\label{eq:MSbarFiniteRemainder}
\mathcal{F}^{A}_{s,b}(a_s, m_t,\mu)
&=& I_{q\bar{q}}\, \mathbf{F}^{A}_{s,b}(a_s, m_t,\mu) = I_{q\bar{q}}\, Z_{q\bar{q}}\, \mathcal{F}^{'A}_{s,b}(a_s, m_t,\mu) \,,
\end{eqnarray} 
where the transformation factor $I_{q\bar{q}}\, Z_{q\bar{q}}$ is free of poles and one can take its 4-dimensional expression.  
To be more specific, to the perturbative order needed here, it reads in the 4-dimensional limit 
\begin{eqnarray}
\label{eq:transformfactor}
I_{q\bar{q}}\, Z_{q\bar{q}}
&=& 1 + a_s \, C_F\, \big(-\ln^2\frac{\mu^2}{-s-i0^+} - 3 \ln \frac{\mu^2}{-s-i0^+} + \frac{\pi^2}{6}  \big)\,+\,\mathcal{O}(a_s^2)\,,
\end{eqnarray} 
where the $a_s$-coefficient is the $\mathcal{O}(\epsilon^0)$ term of the $a_s$-coefficient of $I_{q\bar{q}}$ in eq.(\ref{eq:Iqq}).
The same transformation factor holds for $\mathcal{F}^{A}_{s,t}(a_s, m_t,\mu)$ as well. 
While the RG equations of these $\MSbar$ finite remainders would develop additional terms in the r.h.s. compared to those of eq.~(\ref{eq:FRsRGE}), due to the non-zero anomalous dimension of $Z_{q\bar{q}}$.
Because of this, below we would only analyse the RG equations~(\ref{eq:FRsRGE}) for the finite remainders defined in eq.(\ref{eq:FiniteRemainder}) in 4 dimensions, while switching to the respective $\MSbar$ finite remainders is straightforward as explained in eq.~(\ref{eq:MSbarFiniteRemainder}).

\section{Results for the finite remainders}
\label{sec:res}

With all necessary ingredients ready, we present in this section our numerical results for the finite remainders of singlet quark FFs with exact top mass dependence over the full kinematic range from the low-energy (large-$m_t$) limit $s/m^2_t \rightarrow 0$ to the high-energy (small-$m_t$) limit $s/m^2_t \rightarrow \infty$.

First, we observe that with our bare results for the axial singlet FFs, all poles in $\epsilon$ cancel after performing UV renormalization and IR subtraction as explained in the previous section, which itself serves as a welcome check.
The parts of the resulting finite remainders featuring explicit logarithms in $\mu$ can be revealed via solving the RG equations~(\ref{eq:FRsRGE}) perturbatively in $a_s$. 
In particular, the $\mu$ logarithms in the three-loop coefficients of the finite remainders are entirely determined from the well-tested lower-order results.
Our 2-loop results are cross-checked with those in ref.~\cite{Bernreuther:2005rw} and agreement is found.
We therefore show below the numerical results for the 3-loop singlet FFs with $\mu$ fixed at $s$ where the purely massless contributions evaluate to constants, serving as convenient reference points.
\begin{figure}[htbp]
\begin{center}
\includegraphics[width=0.8\textwidth]{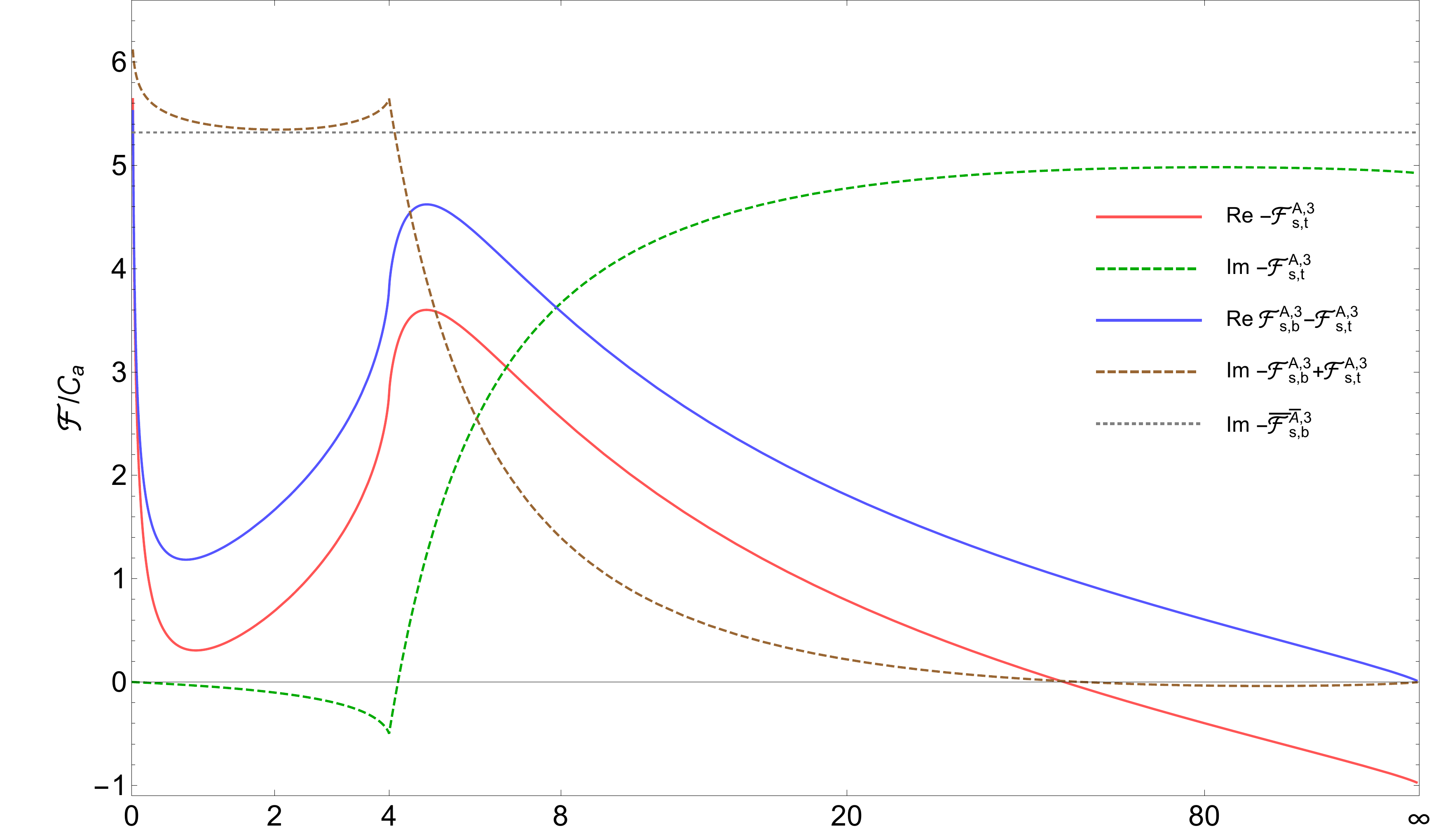}
\caption{The exact result for the 3-loop finite remainders defined in eq.(\ref{eq:FiniteRemainder}), plotted as function of $x = s/m^2_t$, with values normalized w.r.t the real part of the 5-flavor massless result $\mathcal{C}_a \approx 175.218$~\cite{Gehrmann:2021ahy}.}
\label{fig:FF3Lexact_AFF}
\end{center}
\end{figure}
Fig.~\ref{fig:FF3Lexact_AFF} contains our results for the 3-loop finite remainders defined in eq.(\ref{eq:FiniteRemainder}), plotted as function of $x = s/m^2_t$, with values normalized w.r.t the real part of the 5-flavor massless result $\mathcal{C}_a$ which is a constant approximately $175.218$~\cite{Gehrmann:2021ahy}.
In terms of our notations introduced below in eq.(\ref{eq:nonDCsBdiagrams}), it reads 
$\mathcal{C}_a = \mathrm{Re}\big[\bar{\mathcal{F}}^{\bar{A},3}_{s,b}(\mu^2=s,n_l=5)\big]$.
This plot is made from a sample of these functions at about $2 \times 10^5$ points all evaluated with very high precision.
In the high energy limit $x \rightarrow \infty$, one expects the asymptotic relation $\mathcal{F}^{A}_{s,t}(a_s, x) \rightarrow \mathcal{F}^{A}_{s,b}(a_s, x)$, corresponding to the well known result that the total singlet contribution to axial quark FFs vanishes with 6 massless quarks. 
This is demonstrated in fig.~\ref{fig:FF3Lexact_AFF} by both the blue curve representing the (normalized) real part, and the dashed brown curve representing the (normalized and sign-flipped) imaginary part, approaching 0 at $x \rightarrow \infty$. 
This serves as a strong check of our result for $\mathcal{F}^{A,3}_{s,t}(x)$, given the correctness of the analytically-known purely massless result.
The dotted gray line represents the ratio between  
$-\mathrm{Im}\big[\bar{\mathcal{F}}^{\bar{A},3}_{s,b}(x,n_l=5)\big] \approx 931.771$~\cite{Gehrmann:2021ahy} and  $\mathcal{C}_a$, which is about $5.318$, included for reference purpose.
The dashed green curve, standing for $-\mathrm{Im}\big[\mathcal{F}^{A,3}_{s,t}(x)\big]/\mathcal{C}_a$, does not overlap with this in the high energy limit, and this is simply due to the fact that it has effectively $n_f=6$ massless quark loops in the gluon self-energy insertion while the reference line has $n_l=5$ by choice.
Because of the same reason, the red curve does not really arrive at -1, deviating by about $1.5 \%$, albeit almost invisible from the plot.

As one lowers the energy down around the top pair threshold at $x = 4$, one observes the typical behavior due to the Coulomb effect in the real and imaginary part of the top-loop contribution $\mathcal{F}^{A,3}_{s,t}(x)$, of which the former still varies smoothly while the later experiences a (non-smooth) sharp turn.
There is no actual divergence observed in this 3-loop result, because we have here at most one virtual gluon exchange between the virtual top pair.

\begin{figure}[htbp]
\begin{center}
\includegraphics[width=0.8\textwidth]{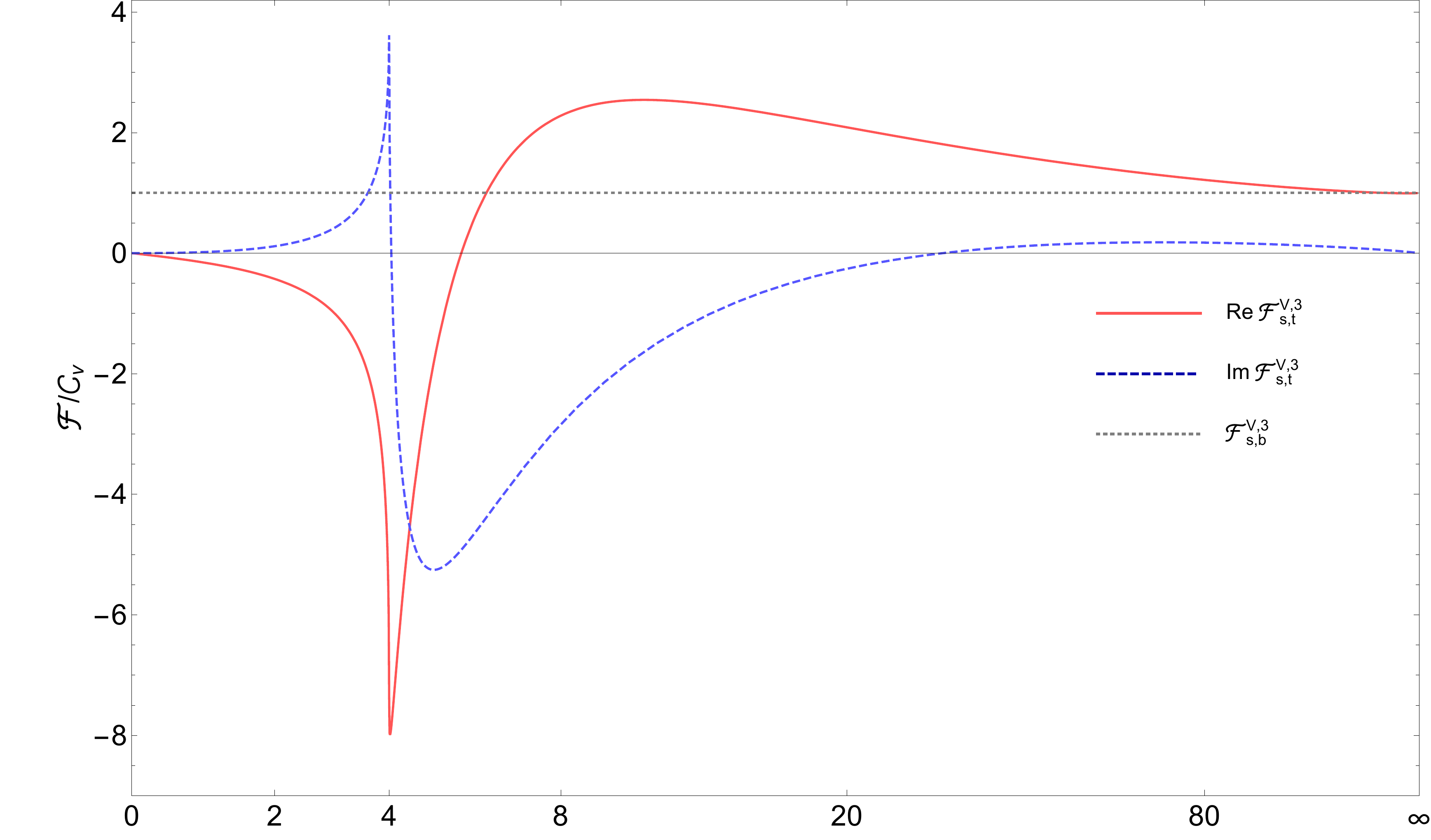}
\caption{
The exact result for $\mathcal{F}^{V,3}_{s,t}(x)$ plotted as function of $x = s/m^2_t$, normalized w.r.t the massless result $\mathcal{C}_v \equiv  \mathcal{F}^{V,3}_{s,b} \approx  -5.94$~\cite{Moch:2005id,Baikov:2009bg,Gehrmann:2010ue}.}
\label{fig:FF3Lexact_VFF}
\end{center}
\end{figure}

Below the top pair threshold $x < 4$, one enters the domain where in principle the large top mass expansion can approximate the full result well, given power corrections included to sufficiently high orders. 
What is special here for the axial FF is that the top quark loop effect does not decouple in the large top mass or low energy limit~\cite{Collins:1978wz,Chetyrkin:1993jm,Chetyrkin:1993ug,Larin:1993ju,Larin:1994va}.
This is reflected in the plot by both the red and blue curve soaring up as $x \rightarrow 0$.~\footnote{The enhancement is only powers of logarithms in $s/m^2_t$, however, the x-axis is not linear in $s/m^2_t$ in order to include the $s/m^2_t \rightarrow \infty$ limit in the plot.} 
This is in contrast to the low energy behavior of the top singlet contribution to the vector counterpart shown in fig.~\ref{fig:FF3Lexact_VFF}, where they are clearly power-suppressed.
We note, however, the imaginary part of the $\mathcal{F}^{A,3}_{s,t}(x)$ is power-suppressed in the limit $x \rightarrow 0$, while that of $\mathcal{F}^{A,3}_{s,b}(x)$ still features a logarithmic enhancement.

Before we dive into the examination of the quality of the large top mass expansion in this region, let us remark that given the typical size of the normalized axial FF values plotted in fig.~\ref{fig:FF3Lexact_AFF} over a wide range, without incorporating the coherent top-loop diagrams, the result for the singlet contribution to the axial part of the quark FF would be, in general, completely off. 
Furthermore, their role in getting the proper $\mu$ dependence of the total singlet contribution, as well as stabilizing it in a truncated perturbative result, is evident from the RG equations discussed in the previous section. 
~\\

We now zoom into the low energy region, and examine the quality of the large top mass expansion. 
In this region it is more sensible to renormalize the perturbative coupling $a_s$ such that the top quark effect in the gluon's self-energy correction is decoupled. 
This is implemented in the form of a finite decoupling renormalization of $a_s = \zeta_{\alpha} \, \bar{a}_s$ with $\zeta_{\alpha} = 1 + \bar{a}_s\, \frac{2}{3}\, \ln \frac{\mu^2}{m^2_t} \, + \,\mathcal{O}(\bar{a}_s^2)$.
Re-expand the finite remainders in powers of $\bar{a}_s$, the perturbative coupling in effective QCD with $n_l = n_f - 1$ massless quark flavors, and one has
\begin{eqnarray}
\label{eq:FiniteRemainderDCas}
\bar{\mathcal{F}}^{A}_{s,b}(\bar{a}_s, m_t,\mu) &=& \mathcal{F}^{A}_{s,b}(a_s= \zeta_{\alpha} \, \bar{a}_s, m_t,\mu)  \nonumber\\ 
&=& \bar{a}_s^2\, \bar{\mathcal{F}}^{A,2}_{s,b}(\mu) \,+\, \bar{a}_s^3\, \bar{\mathcal{F}}^{A,3}_{s,b}(m_t,\mu) 
\,+\, \mathcal{O}(\bar{a}_s^4) 
\,, \nonumber\\
\bar{\mathcal{F}}^{A}_{s,t}(\bar{a}_s, m_t,\mu) &=& \mathcal{F}^{A}_{s,t}(a_s = \zeta_{\alpha} \, \bar{a}_s, m_t,\mu) \nonumber\\ 
&=& \bar{a}_s^2\, \bar{\mathcal{F}}^{A,2}_{s,t}(m_t, \mu) \,+\, \bar{a}_s^3\, \bar{\mathcal{F}}^{A,3}_{s,t}(m_t,\mu) \,+\, \mathcal{O}(\bar{a}_s^4) 
 \,,
\end{eqnarray} 
where we used symbols with a bar to denote the $a_s$-decoupled counterparts of the quantities appearing in previous equations.
Setting $\mu^2 = s$, the perturbative expansion coefficients in eq.(\ref{eq:FiniteRemainderDCas}) become univariate functions of $x = s/m_t^2$, and each can be further expanded as a power-log asymptotic series in $x$ in the limit $x \rightarrow 0$.
To be specific, the $\bar{a}^3_s$ coefficient admits the following expansion ansatz 
\begin{eqnarray}
\label{eq:FFexpanded}
\bar{\mathcal{F}}^{A,3}_{s}(x) \equiv \bar{\mathcal{F}}^{A,3}_{s,b}(x) - \bar{\mathcal{F}}^{A,3}_{s,t}(x) &=& \sum_{n=0}^{\infty}\sum_{m=\underline{m}_n}^{\overline{m}_n} c_{n,m}\, x^n\,\ln^m x
\end{eqnarray} 
where the integer $m$ is bounded within a $n$-dependent range, and the power $n$ is truncated up to certain value in practical calculation.
In particular, for the leading power approximation of $\bar{\mathcal{F}}^{A,3}_{s}(x)$ to be plotted below, one keeps only the terms without powers in $x$.
\begin{figure}[htbp]
\begin{center}
\includegraphics[width=0.8\textwidth]{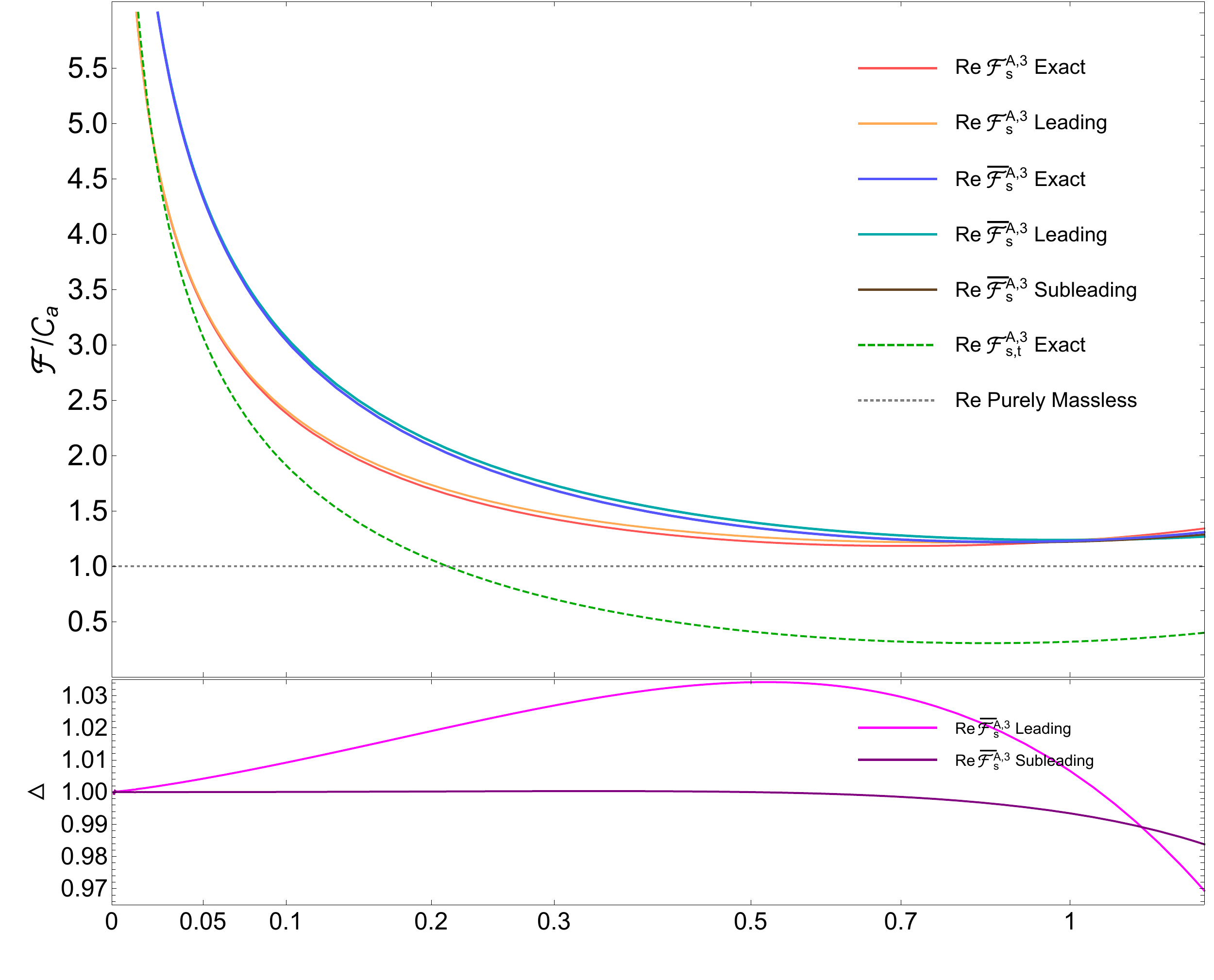}
\caption{
Comparison of the real part of the exact result for $\bar{\mathcal{F}}^{A,3}_{s}(x)$ defined in eq.(\ref{eq:FFexpanded}) with its leading and subleading large mass approximation, as function of $x = s/m_t^2$ in the range $(0, 1.33)$.
The FFs are normalized w.r.t the real part of the 5-flavor massless result $\mathcal{C}_a \approx 175.218$~\cite{Gehrmann:2021ahy}.
The curves representing the exact result for $\mathcal{F}^{A,3}_{s}(x)$ introduced below eq.(\ref{eq:FRsRGE}) and its leading large-mass approximation are included for reference.
The lower panel shows the ratios of the leading and subleading approximation to the exact result.
A similar plot for the imaginary part is given in fig.~\ref{fig:FF3LLER_Im}.}
\label{fig:FF3LLER_Re}
\end{center}
\end{figure}
In fig.~\ref{fig:FF3LLER_Re} and fig.~\ref{fig:FF3LLER_Im}, we show the comparison of the exact result for $\bar{\mathcal{F}}^{A,3}_{s}(x)$ with its leading and subleading large-mass approximation as function of $x$ in the range $(0, 1.33)$ (corresponding to $\sqrt{s} \in (0, 200)$~GeV if one sets $m_t = 173$~GeV), respectively for the real and imaginary part. 
\begin{figure}[htbp]
\begin{center}
\includegraphics[width=0.8\textwidth]{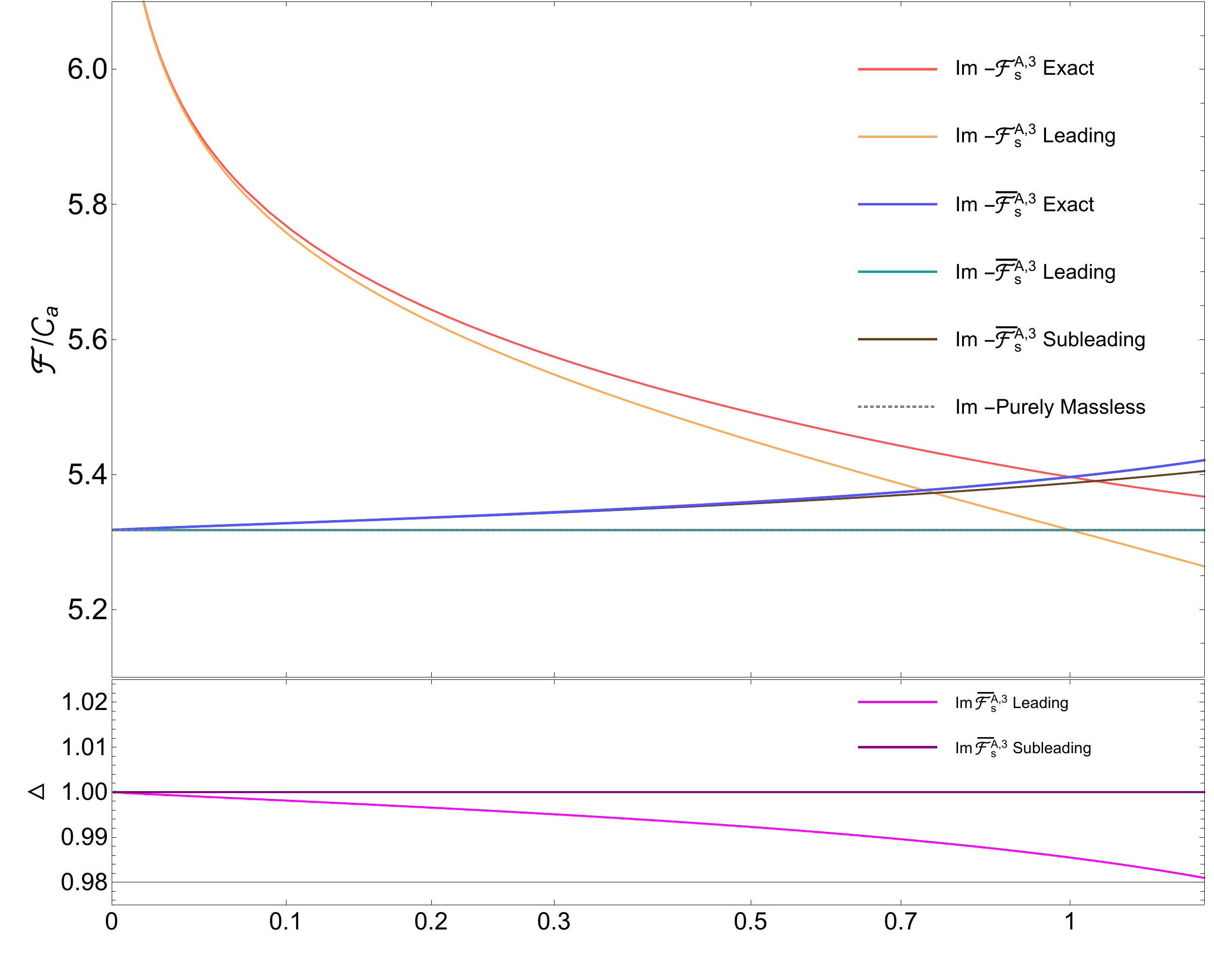}
\caption{
Comparison of the imaginary part of the exact result for $\bar{\mathcal{F}}^{A,3}_{s}(x)$ with its leading and subleading large-mass approximation, as function of $x = s/m_t^2$ in the range $(0, 1.33)$, plotted in the similar way as in fig.~\ref{fig:FF3LLER_Re}.}
\label{fig:FF3LLER_Im}
\end{center}
\end{figure}
As clearly demonstrated in the lower panels, the accuracy of the leading large-mass approximation within this range is already quite good for $\bar{\mathcal{F}}^{A,3}_{s}(x)$, deviating from the exact result by at most $3 \%$.
Including the subleading power correction $\mathcal{O}(x^1)$ further reduces the deviation to be below $1\%$ in most of the range covered, hardly visible in the plot.
The curves representing our exact result for the $a^3_s$-coefficient $\mathcal{F}^{A,3}_{s}(x)$ and its leading large-mass approximation are included for reference.
In particular, one sees that the logarithmic enhancement in its imaginary part is removed by the $a_s$-decoupling in eq.(\ref{eq:FiniteRemainderDCas}).
Consequently in the leading power approximation $\mathrm{Im}\big[\bar{\mathcal{F}}^{A,3}_{s}(x)\big]$ is just a constant, given precisely by the purely massless result, leading to the overlap between the solid cyan line overlaps with the dashed gray line.

In the supplemental material associated with this article, we attach the expanded result for the finite remainder $\bar{\mathcal{F}}^{A,3}_{s}(x)$ truncated to  $\mathcal{O}(x^{10})$, which is very compact.~\footnote{The exact numerical  result is available upon request.}
For the x-range covered in the plot, the difference between this expanded and the exact result is below $10^{-5}$.
It is sufficient to approximate the full result at the level of one per-mille up to the point $x = 3$, corresponding to $\sqrt{s} \approx 300$~GeV with $m_t = 173$~GeV, and below $3 \%$ up to $x = 3.76$.
As discussed at the end of the section~\ref{sec:uvir}, it is straightforward to transform this finite remainder defined under the convention of eq.(\ref{eq:FiniteRemainder}) to others in different IR subtraction schemes that one may use in physical applications.

\section{The Wilson coefficient in the low-energy effective Lagrangian}
\label{sec:wcv}

As mentioned already in the introduction, it is well known~\cite{Collins:1978wz,Chetyrkin:1993jm,Chetyrkin:1993ug,Larin:1993ju,Larin:1994va} that the effect of top quark loops in axial FFs does not decouple in the large top mass or low energy limit due to the presence of the axial-anomaly type diagrams.
In the large top mass limit, the appearance of these non-decoupled mass logarithms in the (IR-subtracted) finite remainders of axial FFs are accompanied by the dependence on the renormalization scale $\mu$ as described by the RG equations of individual singlet contributions discussed in section~\ref{sec:uvir}.
Therefore, the top contribution $\mathcal{F}^{A}_{s,t}(a_s, m_t,\mu)$ can not completely decouple in the naive sense in the large top mass or low energy limit, because the ``massless'' contribution $\mathcal{F}^{A}_{s,b}(a_s, m_t,\mu)$ still has a non-zero anomalous dimension to be compensated such that the total anomaly-free result has the expected $\mu$ dependence~\cite{Collins:1978wz,Chetyrkin:1993jm,Chetyrkin:1993ug,Larin:1993ju,Larin:1994va}. 
Once combined together in the form as in eq.(\ref{eq:FFsRGEphys}), the explicit remaining $\mu$ dependence is again dictated by just the $\MSbar$ renormalization of the $a_s$.
However, the non-power-suppressed $m_t$-logarithms would not drop and remain in the form of $\ln \frac{s}{m^2_t}$ in the total result.

\begin{figure}[htbp]
\begin{center}
\includegraphics[scale=0.5]{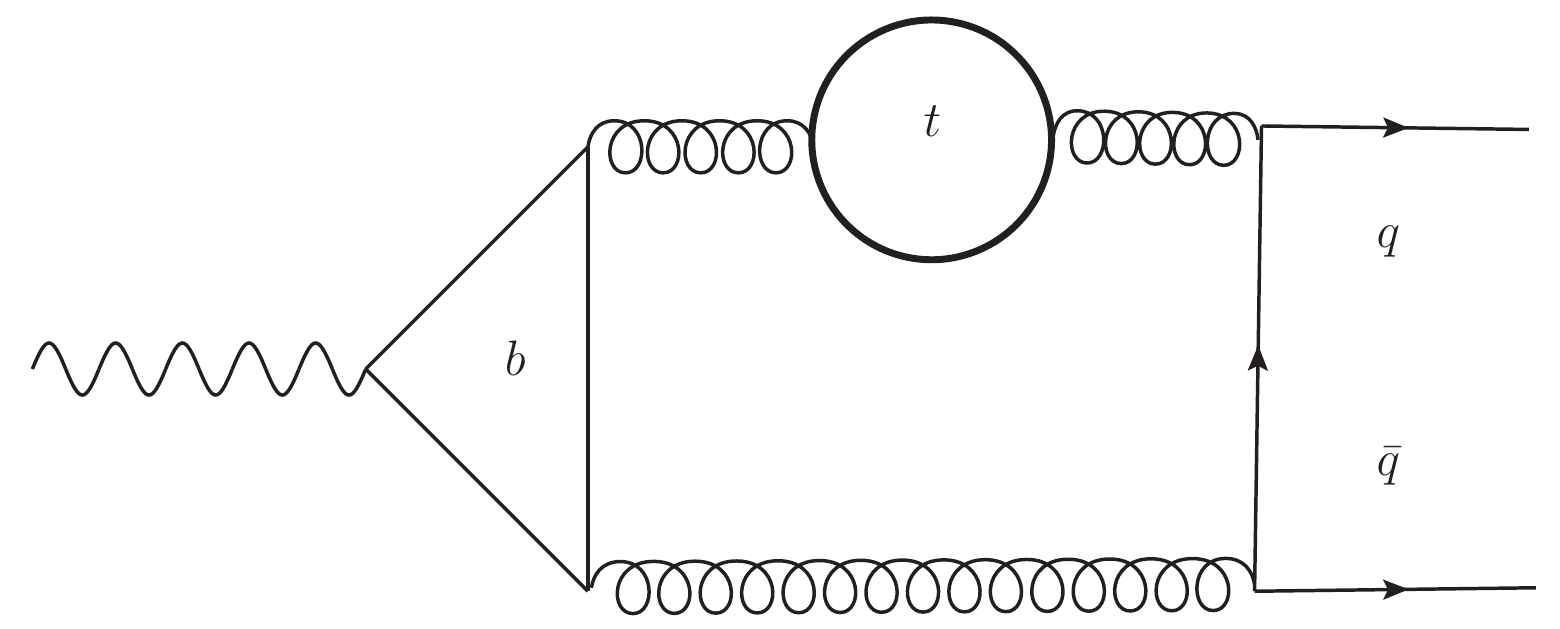}
\caption{A representative diagram contributing to $\mathit{F}^{A}_{s,b}$ with top loop insertion.}
\label{fig:singlet3L_topbub}
\end{center}
\end{figure}

Still, it is quite striking to observe that even the four $m_t$-dependent 3-loop Feynman diagrams contributing to $\bar{\mathcal{F}}^{A,3}_{s,b}(m_t,\mu)$, all with top  loop insertion through the gluon self-energy correction with an example drawn in fig.~\ref{fig:singlet3L_topbub}, generate non-power-suppressed $m_t$-logarithms beyond those that would be removed by the usual decoupling renormalization of $a_s$.
To be more specific, 
\begin{eqnarray}
\label{eq:nonDCsBdiagrams}
\bar{\mathcal{F}}^{A,3}_{s,b}(m_t,\mu)\Big|_{m_t \rightarrow \infty}  
&=&  \bar{\mathcal{F}}^{\bar{A},3}_{s,b}(\mu) \,+\, \bar{\mathcal{F}}^{A_{\text{{\tiny nDC}}},3}_{s,b}(m_t,\mu)\,,\nonumber\\
&=& \bar{\mathcal{F}}^{\bar{A},3}_{s,b}(\mu)  
-\frac{85}{9} C_F + \frac{4}{3} C_F L_{\mu} - \frac{1}{4} C_F L_{\mu}^2
\,+\,\mathcal{O}(1/m_t^2)\, 
\end{eqnarray} 
where $L_{\mu} \equiv \ln \frac{\mu^2}{m^2_t}$. 
Similar non-decoupling terms were determined in refs.~\cite{Chetyrkin:1993ug,Larin:1994va} for the singlet contribution to the inclusive Z boson decay rate in the large top mass limit.
We note that only after explicitly decoupling the top loop effect from the $a_s$ renormalization, as done in eq.(\ref{eq:FiniteRemainderDCas}), the remaining non-decoupled $m_t$-logarithms collected in $\bar{\mathcal{F}}^{A_{\text{{\tiny nDC}}},3}_{s,b}(m_t,\mu)$ then appear solely in terms of $\ln \frac{\mu^2}{m^2_t}$ (i.e.~fully matching with the $\mu$ dependence therein).
We note that $\bar{\mathcal{F}}^{\bar{A}}_{s,b}(\mu)$ is the result in effective top-less QCD with $n_l = n_f -1$ massless quarks, free of the top mass from the outset.
The non-decoupling $m_t$-logarithms observed in $\bar{\mathcal{F}}^{A_{\text{{\tiny nDC}}},3}_{s,b}(m_t,\mu)$ are checked to obey the expanded form of the RG equation~(\ref{eq:FRsRGE}) truncated to the perturbative order in question.
~\\

For the axial FF of the flavor-$q$ quark, a coefficient function $C_w(\bar{a}_s,\mu/m_t)$ is introduced to encode all the remaining non-decoupling $m_t$-logarithms appearing in the total ``physical'' (or non-anomalous) combination $\mathcal{F}^{A}_{s,b}(a_s, m_t,\mu) - \mathcal{F}^{A}_{s,t}(a_s, m_t,\mu)$ in the following way:
\begin{eqnarray}
\label{eq:Cwdef}
\mathcal{F}^{A}_{s,b}(a_s, m_t,\mu) - \mathcal{F}^{A}_{s,t}(a_s, m_t,\mu) \Big|_{{\tiny m_t \rightarrow \infty}} &=& \bar{\mathcal{F}}^{\bar{A}}_{s,b}(\bar{a}_s,\mu) + \bar{\mathcal{F}}^{A_{\text{{\tiny nDC}}}}_{s,b}(\bar{a}_s,m_t,\mu) - \bar{\mathcal{F}}^{A}_{s,t}(\bar{a}_s, m_t,\mu)\Big|_{{\tiny m_t \rightarrow \infty}}\nonumber\\
&=& \bar{\mathcal{F}}^{\bar{A}}_{s,b}(\bar{a}_s, \mu) \nonumber\\
&-& C_w(\bar{a}_s,\mu/m_t)\, \Big(\bar{\mathcal{F}}^{A}_{ns}(\bar{a}_s, \mu) + \sum_{i=1}^{n_l} \bar{\mathcal{F}}^{\bar{A}}_{s,i}(\bar{a}_s, \mu) \Big)\,\nonumber\\
&+& \,\mathcal{O}(1/m_t^2)\, ,
\end{eqnarray}
where only the remaining non-decoupling $m_t$-logarithms after performing $a_s$-decoupling are absorbed into $C_w(\bar{a}_s,\mu/m_t)$, indicated by the expansion done in powers of $\bar{a}_s$.
By the virtue of RG invariance of the l.h.s. of eq.(\ref{eq:Cwdef}), one can then derive the following RG equation of $C_w(\bar{a}_s,\mu/m_t)$:  
\begin{eqnarray}
\label{eq:CwRGE}
\mu^2\frac{\mathrm{d}}{\mathrm{d} \mu^2} C_w(\bar{a}_s, \mu/m_t) 
&=& \mu^2\frac{\partial}{\partial \mu^2} C_w(\bar{a}_s, \mu/m_t)  \,+\, \bar{\beta}\, a_s\frac{\partial}{\partial a_s} C_w(\bar{a}_s, \mu/m_t)  \nonumber\\
&=& \bar{\gamma}_s - \bar{\gamma}_S \,C_w(\bar{a}_s, \mu/m_t)\,,
\end{eqnarray}
with the aid of   
\begin{eqnarray}
\label{eq:sffRGE}
\mu^2\frac{\mathrm{d}}{\mathrm{d} \mu^2} \bar{\mathcal{F}}^{\bar{A}}_{s,q}(\bar{a}_s, \mu) 
&=& \bar{\gamma}_s \, \Big( \bar{\mathcal{F}}^{A}_{ns}(a_s,\mu) + \sum_{i=1}^{n_l} \bar{\mathcal{F}}^{\bar{A}}_{s,i}(\bar{a}_s, \mu) \Big) \,,
\end{eqnarray}
where $\bar{\beta}$, $\bar{\gamma}_s$ and $\bar{\gamma}_S = n_l \, \bar{\gamma}_s$ are the counterparts of $\beta$, $\gamma_{s}$, and $\gamma_{S}$ in effective QCD with $n_l=5$ massless quarks in 4 dimensions.
Resummation of the non-decoupling $m_t$-logarithms absorbed in $C_w(\bar{a}_s,\mu/m_t)$ can be performed by solving the RG eq.~(\ref{eq:CwRGE}) with the 4-dimensional $\bar{\gamma}_s$ (and $\bar{\gamma}_S$).
We note that the particular form of eq.(\ref{eq:CwRGE}) holds for $C_w(\bar{a}_s,\mu/m_t)$ defined with the aforementioned non-decoupling $m_t$-logarithms in $\mathcal{F}^{A}_{s,b}(a_s, m_t,\mu)$ included, rather than just from the $a_t$-dependent contribution.
For instance, only after this combination, will one observe the number of fermions appearing in $\bar{\gamma}_s$ (and $\bar{\gamma}_S$) becoming $n_l$, as checked to $\mathcal{O}(\bar{a}_s^3)$ explicitly below.
Expanded perturbatively up to order $\bar{a}_s^3$ considered in present work, eq.(\ref{eq:CwRGE}) reduces to a simpler form
\begin{eqnarray}
\label{eq:CwRGEas3}
\mu^2\frac{\mathrm{d}}{\mathrm{d} \mu^2} C_w(\bar{a}_s, \mu/m_t) &=& \bar{\gamma}_s + \mathcal{O}(\bar{a}_s^4)\,.
\end{eqnarray}

With our results for the finite remainders of the FFs, presented in the previous section determined in a non-$\MSbar$ renormalization scheme, we can extract the result for the coefficient $C_w(\bar{a}_s,\mu/m_t)$ defined in eq.(\ref{eq:CwRGE}).
It reads 
\begin{eqnarray}
\label{eq:CwRes}
C_w(\bar{a}_s, \mu/m_t) &=& \bar{a}_s^2 \, \Big( -6 C_F L_{\mu} + 3 C_F \Big) \nonumber\\
&+& \bar{a}_s^3 \, \Big(
L_{\mu}^2 \,\big( -22 C_A C_F + 4 C_F n_l \big)
+ L_{\mu} \, \big( -\frac{76}{3} C_A C_F + 18 C_F^2 - \frac{8}{3} C_F n_l \big) \nonumber\\
&+& C_A C_F \,\big(84 \zeta_3 - \frac{1649}{18}\big) + C_F^2\, \big(-72 \zeta_3 - \frac{51}{2}\big) 
+ C_F \, \big(\frac{187}{9} n_l + \frac{164}{9}\big) 
\Big)
\nonumber\\
&+& \mathcal{O}(\bar{a}_s^4)\,.
\end{eqnarray}
We subsequently checked that its $\mu$ dependence, hence the structure of the non-decoupling $m_t$-logarithms, satisfies eq.(\ref{eq:CwRGE}) or rather its expanded form eq.(\ref{eq:CwRGEas3}) truncated to $\mathcal{O}(\bar{a}_s^3)$, which serves as a check of the calculation.
Note that the $\mu$-logarithm-independent terms of $C_w(\bar{a}_s,\mu/m_t)$ in eq.(\ref{eq:CwRes}) can not be determined by the RG eq.(\ref{eq:CwRGE}) alone (with given $\bar{\gamma}_s$ of course), but have to be extracted from explicit calculations.
In general, these constant terms depend on the renormalization scheme in use.
In fact, before being combined together, the coefficients of the remaining non-decoupling $m_t$-logarithms from $a_b$- and $a_t$-dependent parts in eq.(\ref{eq:Cwdef}) are respectively renormalization-scheme dependent starting from 3-loop order, e.g.,~the coefficient of $L_{\mu}$ in eq.(\ref{eq:nonDCsBdiagrams}).
The $\mathcal{O}(\bar{a}_s^2)$ coefficient in eq.(\ref{eq:CwRes}) agrees with refs.~\cite{Chetyrkin:1998mw,Ahmed:2020kme} where it was extracted from explicit calculations of different quantities determined in the same renormalization scheme as used here.
The $\mathcal{O}(\bar{a}_s^3)$ coefficient in eq.(\ref{eq:CwRes}) is found in agreement with the expression given in a recent publication~\cite{Ju:2021lah} where it was composed from the $\overline{\mathrm{MS}}$ result in ref.~\cite{Chetyrkin:1993ug} with the aid of ref.~\cite{Ahmed:2021spj}.
~\\

As usual, the large mass limit result parameterized in eq.(\ref{eq:Cwdef}) can be generated from, or rather encoded by, an effective Lagrangian with a certain set of relevant local composite operators.
To this end, the first point to notice is that the only type of additional operators relevant in the leading large mass approximation for the amplitude in question is the \textit{singlet} axial current operator~\cite{Collins:1978wz,Chetyrkin:1993jm,Chetyrkin:1993ug} as introduced in eq.(\ref{eq:ZScurrent}).
The part with Z boson couplings in the resulting top-less effective Lagrangian with $n_l$ massless quarks, reads, in its renormalized form:
\begin{eqnarray}
\label{eq:Leff}
\delta \mathcal{L}^{R}_{\mathrm{eff}} &=& \text{Z}_{\mu}\, \mu^\epsilon \Big( \bar{Z}_{ns}\,
\sum_{i=1}^{n_l} a_i \,  \bar{\psi}^{B}_{i}  \, \gamma^{\mu}\gamma_5 \, \psi^{B}_i \,+\,
a_b \, \bar{Z}_{s} \, \bar{J}^{\mu}_{S,5} \nonumber\\
\,&+&\, a_t\, C_w(\bar{a}_s, \mu/m_t)\, \big(\bar{Z}_{ns} + n_l\,\bar{Z}_{s} \big) \, \bar{J}^{\mu}_{S,5} \Big)
\end{eqnarray}
where $\bar{J}^{\mu}_{S,5} = \sum_{i=1}^{n_l} \, \bar{\psi}^{B}_{i}  \, \gamma^{\mu}\gamma_5 \, \psi^{B}_i$ with a non-anticommuting $\gamma_5$, 
and $\bar{Z}_{ns}\,, \bar{Z}_{s}$ denote the counterparts of the axial-current renormalization constants (c.f.~eq.(\ref{eq:ZsAMD})) in this effective QCD.
The combination of the non-decoupling $m_t$-logarithms from both $a_b$- and $a_t$-dependent singlet contribution is accounted for by the term in the second line of eq.(\ref{eq:Leff}), appended to the usual $n_l=5$ purely massless QCD.
Only the sum of the second term in the first line and the second line is non-anomalous, as can be checked directly with the aid of RG equations of pieces given previously.\footnote{We note that the overall factor $\mu^\epsilon$ plays no role in the RG eq.(\ref{eq:CwRGE}) for $C_w(\bar{a}_s,\mu/m_t)$ in 4 dimensions.}

We emphasize that the Wilson coefficient appearing in eq.(\ref{eq:CwRGE}) and eq.(\ref{eq:Leff}) is determined in a non-$\MSbar$ renormalization scheme, in particular for the individual axial currents. 
This is intertwined with the particular renormalized form of axial currents appearing in the low energy effective Lagrangian given in eq.(\ref{eq:Leff}).
The $\mu$-logarithm-independent terms of $C_w(\bar{a}_s,\mu/m_t)$ given in eq.(\ref{eq:CwRes}) depend on the renormalization scheme in use.
And the $C_w(\bar{a}_s,\mu/m_t)$ term so-defined is supposed to be used in calculations made with an effective top-less QCD in the standard way where the chiral Ward identity of the $\bar{\psi}^{B}_{b}  \, \gamma^{\mu}\gamma_5 \, \psi^{B}_b$ is properly restored.
In particular, the top-loop induced contribution to singlet axial FFs determined in this work can be directly combined with the result for the bottom contribution $\bar{\mathcal{F}}^{\bar{A}}_{s,b}(\bar{a}_s, \mu)$ derived in ref.~\cite{Gehrmann:2021ahy}.

\section{Conclusion}
\label{sec:conc}

In this work we determined numerically the finite remainders of the singlet contribution to quark FFs with exact top mass dependence over the full kinematic range, both for the axial and the vector part.
We have worked out the renormalization formulae and RG equations for the individual subsets of singlet contributions to the axial FFs, subsequently checked using our explicit results to the perturbative order considered in the present work.

Our numerical investigation shows that without incorporating the coherent top-loop diagrams, the result for the singlet contribution to the axial part of the quark FF would be in general completely off. 
Furthermore, their role in getting the appropriate scale dependence of the total singlet contribution, as well as stabilizing it in a truncated perturbative result, is evident from the RG equations discussed in this article.

A particular low energy effective Lagrangian is composed to encode the leading large top mass approximation of the full result for the axial quark FFs, with the Wilson coefficient defined and extracted in the non-$\MSbar$ renormalization scheme in use.
We note that only with all non-decoupling $m_t$-logarithms from both $a_t$- and $a_b$-dependent singlet diagrams (remaining after $a_s$-decoupling) combined together, will then the Wilson coefficient $C_w(\bar{a}_s,\mu/m_t)$ so-defined obey a simple RG equation in effective QCD as presented in this work.
The accuracy of the leading large top mass approximation, encoded by such a low energy effective Lagrangian, is examined and is shown to be quite good for the 3-loop coefficient of the finite remainder, deviating from the exact result by $\sim 3 \%$ for $\sqrt{s} < 200$ GeV.

The result presented in this work provides one of the missing ingredients needed to push the theoretical predictions of Z-mediated Drell-Yan processes to the third order in QCD coupling, especially at the differential level.

\section*{Acknowledgements}

We thank W.~L. Ju, C.~Duhr and B.~Mistlberger for communication regarding the three-loop Wilson coefficient $C_w$.
This research was supported by the Deutsche Forschungsgemeinschaft (DFG, German Research Foundation) under grants 396021762 - TRR 257 and 400140256 - GRK 2497: The physics of the heaviest particles at the Large Hardon Collider.

\bibliography{SingletFF} 

\bibliographystyle{utphysM}
\end{document}